\documentclass{techmech}
\usepackage{subfigure}

\title{LES of a non-premixed hydrogen flame stabilized by bluff-bodies of various shapes}
\author{%
Agnieszka Wawrzak\textsuperscript{1$\star$}\!, 
Robert Kantoch\textsuperscript{1}\!, and
Artur Tyliszczak\textsuperscript{1}\!, 
}
\affil{\small%
\textsuperscript{1} Czestochowa University of Technology, Faculty of Mechanical Engineering and Computer Science, Al. Armii Krajowej 21, 42-201 Czestochowa, Poland
}
\renewcommand{\corremail}{agnieszka.wawrzak@pcz.pl}
\renewcommand{\tmkeywords}{bluff-body, wavy-wall, LES, hydrogen, non-premixed flame}
\renewcommand{\tmabstract}{Dynamics of flames stabilized downstream of different shape bluff-bodies (cylindrical, square, star) with different wall topologies (flat, wavy) is investigated using large-eddy simulations (LES). A two-stage computational procedure involving the ANSYS software and an in-house academic high-order code is combined to model a flow in the vicinity of the bluff-bodies and a flame formed downstream. The fuel is nitrogen-diluted hydrogen and the oxidizer is hot air in which the fuel auto-ignites. After the ignition, the flame propagates towards the bluff-body surfaces and stabilizes in their vicinity.
It is shown that the flames reflect the bluff-body shape due to large-scale strong vortices induced in the shear layer formed between the main recirculation zone and the oxidizer stream. The influence of the acute corners of the bluff-bodies on the flame dynamics is quantified by analysing instantaneous and time-averaged results. Compared to the classical conical bluff-body the largest differences in the temperature and velocity distributions are observed in the configuration with the square bluff-body. The main recirculation zone is shortened by approximately 15\% and at its end temperature in the axis of the flame is almost 200~K larger. Simultaneously, their fluctuations are slightly larger than in the remaining cases. The influence of the wall topology (flat vs. wavy) in the configuration with the classical conical bluff-body turned out to be very small and it resulted in modifications of the flow and flame structures only in the direct vicinity of the bluff-body surface.}

\begin{document}
\pagestyle{scrheadings} 
\maketitle

\section{Introduction}
A better understanding of mutual interactions between turbulent flow and flame occurring downstream the bluff-body geometries limit the further progress in improvements in the efficiency and safety of various combustion applications (burners, chambers, engines). Moreover, a suitable control of turbulent flames dynamics is required for establishment of low-emissions devices according to current international regulations. Such a control can be achieved by applying passive and/or active flow control techniques. Both provide modulation of the multi-scale mixing processes (enhancement/suppression) by the intensification of interactions between large and small turbulent scales. 

Using the bluff-body as a part of the injection system constitutes the prominent example of the passive flame control. The bluff-body generates recirculation zones improving the mixing and stabilizing the flame position~\citep{docquier2002combustion}. ~\cite{tyliszczak2014cmc} examined a non-premixed flame stabilized in a central recirculation zone produced by a conical/cylindrical bluff-body. They showed that the flame is approaching extinction due to even small changes in simulation parameters. Combination of the passive and active control methods can be applied for optimization of the combustion process in bluff-body burner as it was shown by~\cite{kypraiou2018response}. The researchers performed the experimental studies on the premixed, partially-premixed and non-premixed methane flames showing that the impact of acoustic oscillations on the bluff-body flames is directly related to the fuel injection system. 

The complexity of turbulent mixing and combustion processes raises many important questions that still remain unanswered, despite the great interest of the scientific community in this field. For instance, to what extend does the mixing of the fuel and oxidizer downstream the bluff-body depend on the shape and roughness of the bluff-body? As it was recently demonstrated by~\cite{drozdz2021effective}, the wavy wall, with carefully selected waviness parameters, can effectively enhance the effect of amplitude modulation and hence increase wall shear stress and postpone turbulent separation. 

Concerning the geometrical shaping, interesting findings were recently reported by~\cite{tyliszczak2022numerical}. The authors showed a distinct trend regarding the dynamics of the jets coming out from the shaped orifices. They found a shorter potential core and faster velocity decay for a circular jet compared to a non-circular one, e.g., issuing from a triangular nozzle. This contradicted the common knowledge of the jets emenating from sharp-edged orifices. In general, they are more energetic compared to jets generated by smoothly profiled nozzles, therefore they are characterized by a faster decay of the axial velocity~\citep{mi2010statistical,KubanStempkTyliszczak_Energies_2021}. Nevertheless,~\cite{tyliszczak2022numerical} have shown that the level of control that can be achieved by shaping the nozzle is higher than previously expected.\\

The present paper aims at numerical simulations of non-premixed hydrogen flame stabilized by specially designed bluff-bodies with corrugated surfaces. Computational fluid dynamics (CFD) tools have been widely used for the flame control analyses providing the results related to both the global characteristics as well as concerning deep insight into a flame structure and its dynamics. 

In this work we use the Large Eddy Simulation method and a 'no model' approach for the simulations of the combustion process  where the chemical sources terms are calculated directly using the filtered variables~\citep{duwig2011large}. The response of the flame to different wall topologies (flat or wavy, with the waviness oriented streamwise) and complex shapes (hexagram, square) is thoroughly discussed based on instantaneous and time averaged results.

\section{Modelling}
\subsection{LES Approach} 
In the present study we use two different numerical LES solvers and apply a two-stage computational approach. In the first stage (I), the second order ANSYS Fluent LES solver is involved  to model the flow inside the inlet section of the bluff-body burner. The further calculations (stage II) are performed using a in-house high-order numerical algorithm based on the projection methods~\citep{Tyl16}. It solves the Favre filtered set of the governing equations assuming the low Mach number approximation~\citep{geurts2004elements}:
\begin{equation} \label{eq:con}
\partial_t \bar\rho+\nabla\cdot(\bar\rho\widetilde{\mathbf{u}})=0
\end{equation}
\begin{equation} \label{eq:NS}
\bar{\rho}\partial_t\widetilde{\mathbf{u}} +(\bar{\rho}\widetilde{\mathbf{u}}\cdot\nabla)\widetilde{\mathbf{u}} +\nabla\bar{p}=\nabla\cdot\ (\boldsymbol{\tau}+\boldsymbol{\tau}^{\mathrm{SGS}})
\end{equation}
\begin{equation} \label{eq:spec}
\bar{\rho}\partial_t\widetilde{Y}_{\alpha}+(\bar{\rho}{\widetilde{\mathbf{u}}\cdot \nabla)\widetilde{Y}_{\alpha}}
=\nabla\cdot\left({\bar{\rho}({D}_{\alpha}+{D}^{\mathrm{SGS}}_{\alpha}) {\nabla \widetilde{Y}_{\alpha}}}\right)+ \overline{\dot w}_{\alpha}
\end{equation}
\begin{equation} \label{eq:ent}
\bar{\rho}\partial_t\widetilde{{h}}+(\bar{\rho}{\widetilde{\mathbf{u}}\cdot \nabla)\widetilde{h}}=\nabla\cdot\left({\bar{\rho}({D}+{D}^{\mathrm{SGS}}) {\nabla{h}}}\right)
\end{equation}
\begin{equation}
p_0=\overline{\rho} R \widetilde{T} \label{eq:state}
\end{equation}\\
where the bar and tilde symbols denote filtered quantities, ${u}_i$ are the velocity components, $p$ is the hydrodynamic pressure, $\rho$ is the density and $h$ stands for the total enthalpy. The symbols $p_0$ and $R$ are the thermodynamic pressure and gas constant, respectively. The subscript $\alpha$ is the index of the species $\alpha=1,\dots, \textrm{N-species} $ whereas the variables $Y_\alpha$ represent species mass fractions. An unresolved sub-grid stress tensor, resulting from the filtering of the non-linear advection terms is defined as $\boldsymbol{\tau}^{\mathrm{SGS}}=2\mu_{\mathrm{t}}\mathbf{S}$, where $\mathbf{S}$ is the rate of strain tensor of the resolved velocity field and $\mu_{\mathrm{t}}$ is the sub-grid viscosity  computed according to the model proposed by~\cite{Vreman2004}. 
The sub-grid diffusivities in Eqs.~\ref{eq:spec},~\ref{eq:ent} are computed as ${D}^{\mathrm{SGS}}=\mu_t/(\bar\rho\sigma)$ where $\sigma$ is the turbulent Schmidt/Prandtl number assumed equal to 0.7~\citep{jones2007large}.
\begin{figure}[!ht]
 \begin{center}
  \includegraphics[width=0.85\textwidth]{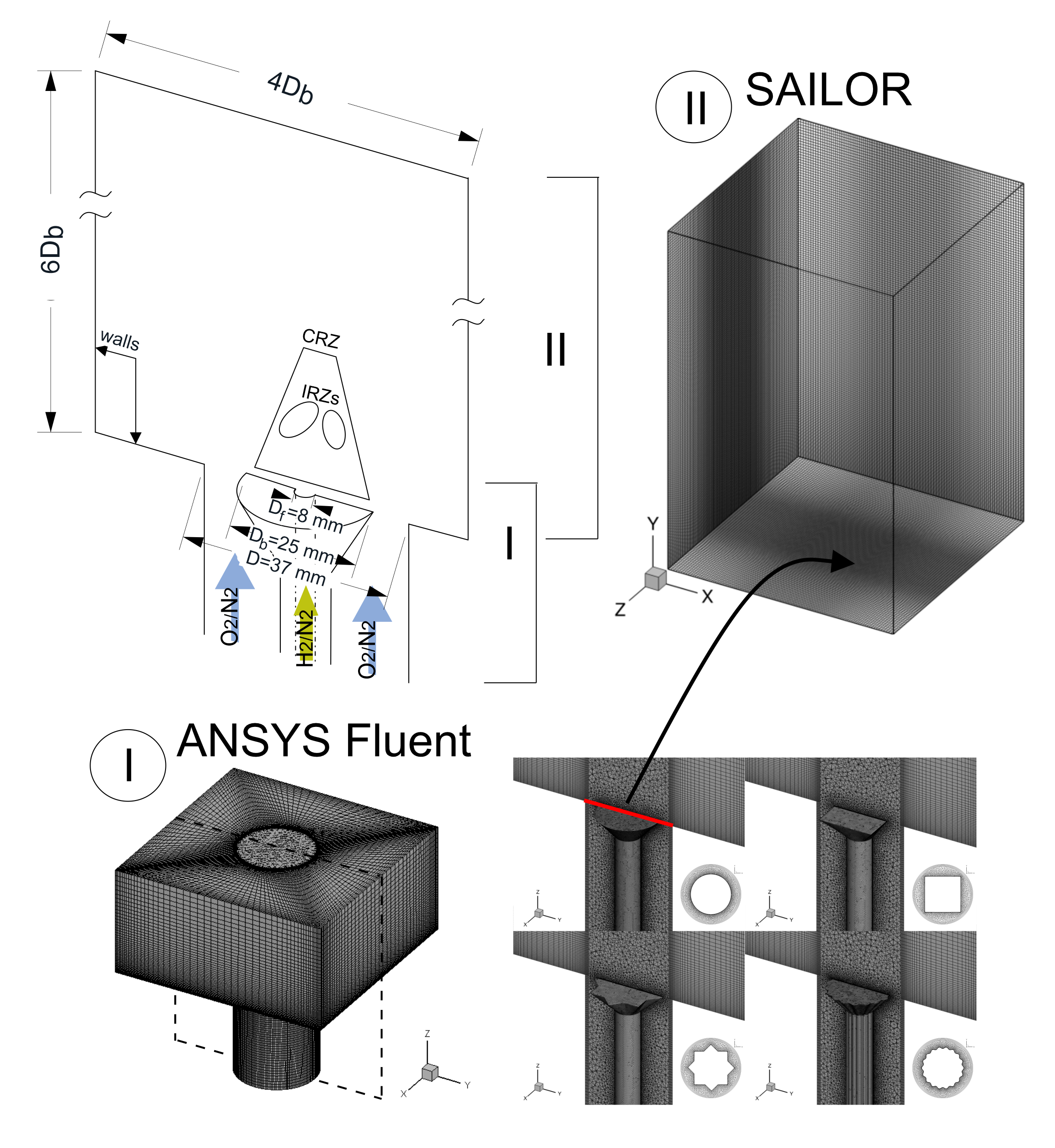}
\end{center}
	\caption{Schematic view on the computational configuration. Bluff-body geometries and computational meshes.}
	\label{fig:BB-config}
	\end{figure}

The chemical sources terms $\overline{\dot w_\alpha}$ in Eq.~\ref{eq:spec} involve the filtered reaction rates of species $\alpha$, which are strongly non-linear functions of the species mass fractions and enthalpy:
\begin{equation} \label{eq:omega}
\overline{\dot w_\alpha(\mathbf{Y},h)}={\dot w_\alpha(\widetilde{\mathbf{Y}},\widetilde{h})}+\mathcal{F}(\widetilde{Y_\alpha {Y_\alpha}^{''}},\widetilde{{Y_\alpha}^{''}T}, \dots)
\end{equation}\\
They are computed applying the 'no model' approach and are calculated directly using the filtered variables $\overline{\dot w_\alpha(\mathbf{Y},h)}={\dot w_\alpha(\widetilde{\mathbf{Y}},\widetilde{h})}$~\citep{duwig2011large}.
\begin{figure}[!ht]
 \begin{center}
  \includegraphics[width=0.23\textwidth]{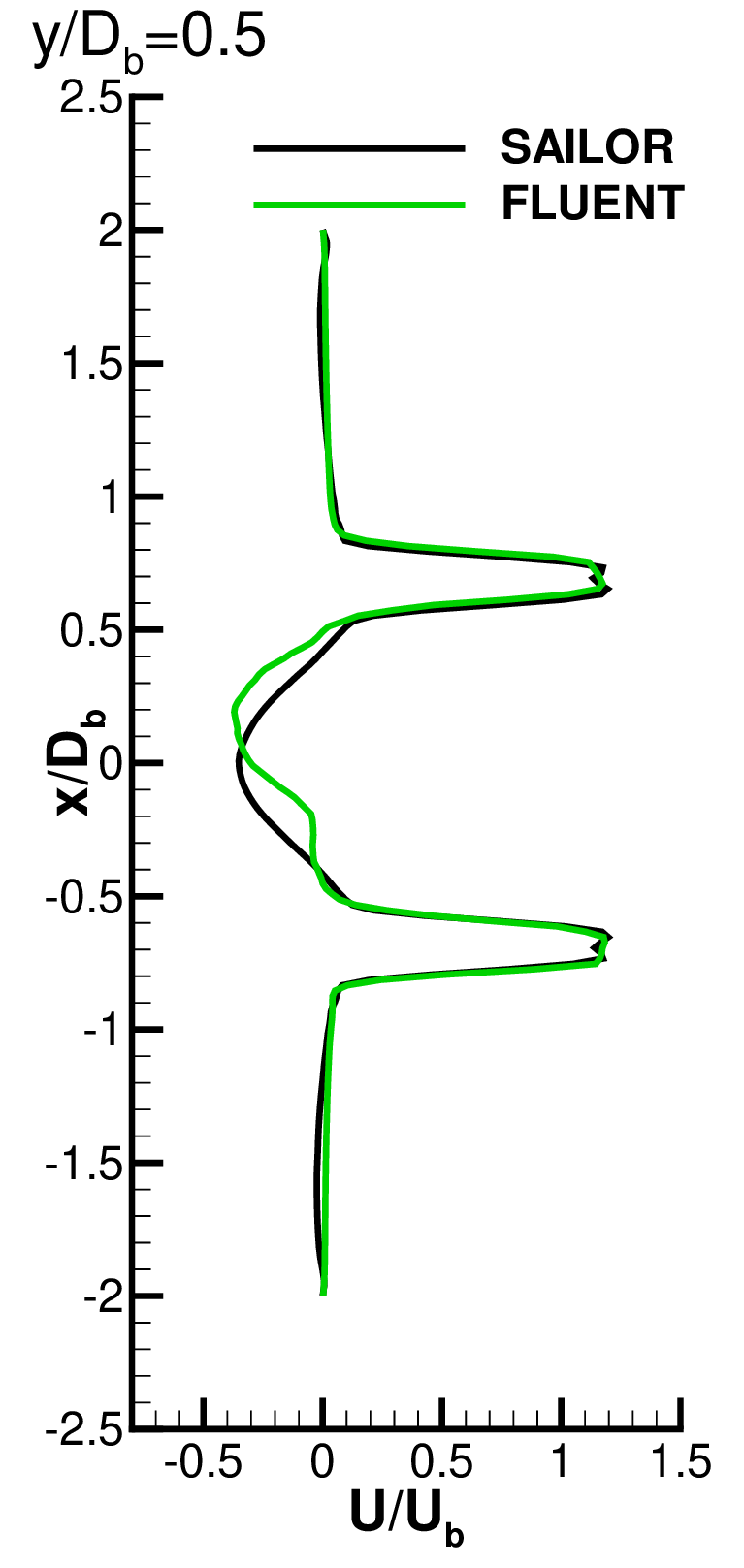}
  \includegraphics[width=0.23\textwidth]{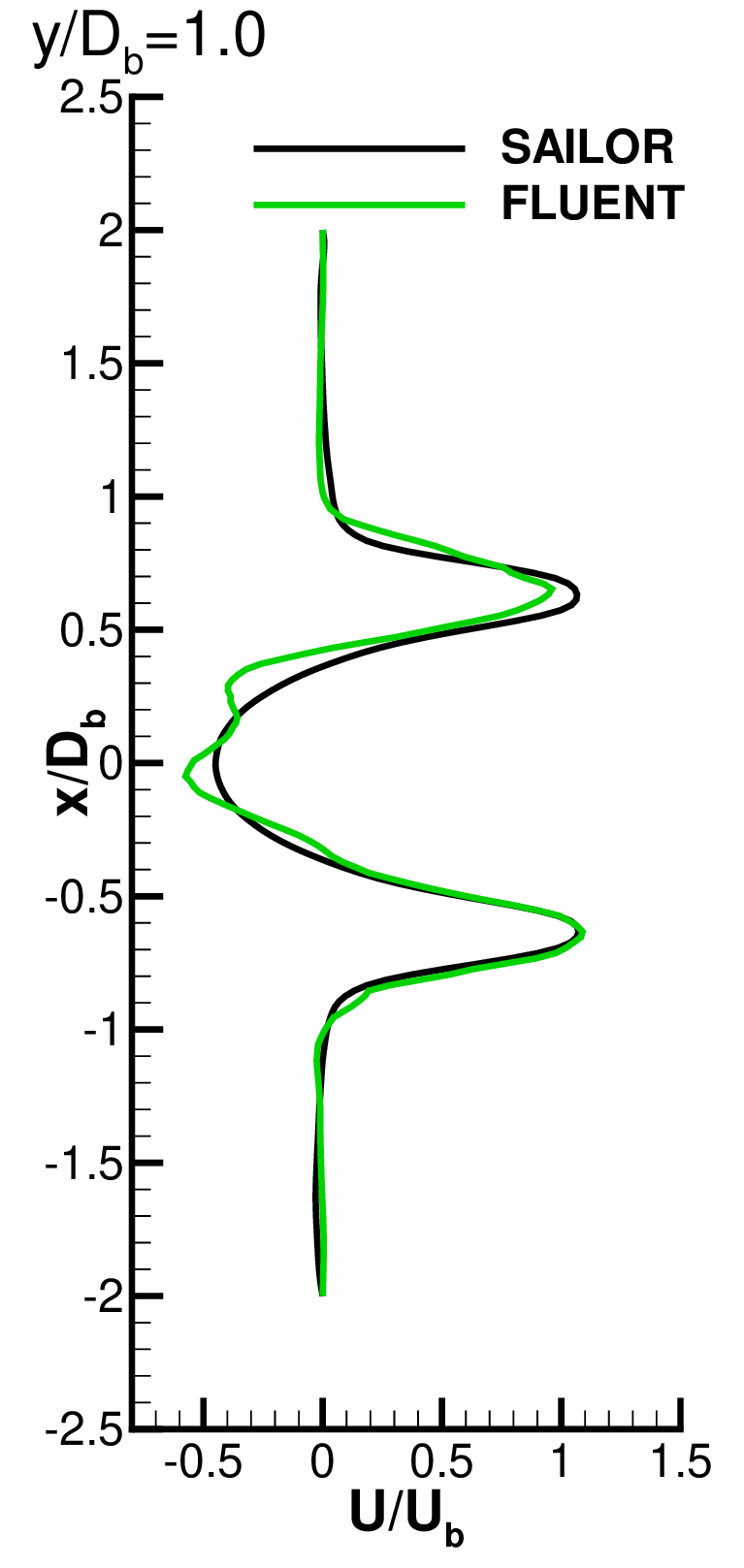}
  \includegraphics[width=0.23\textwidth]{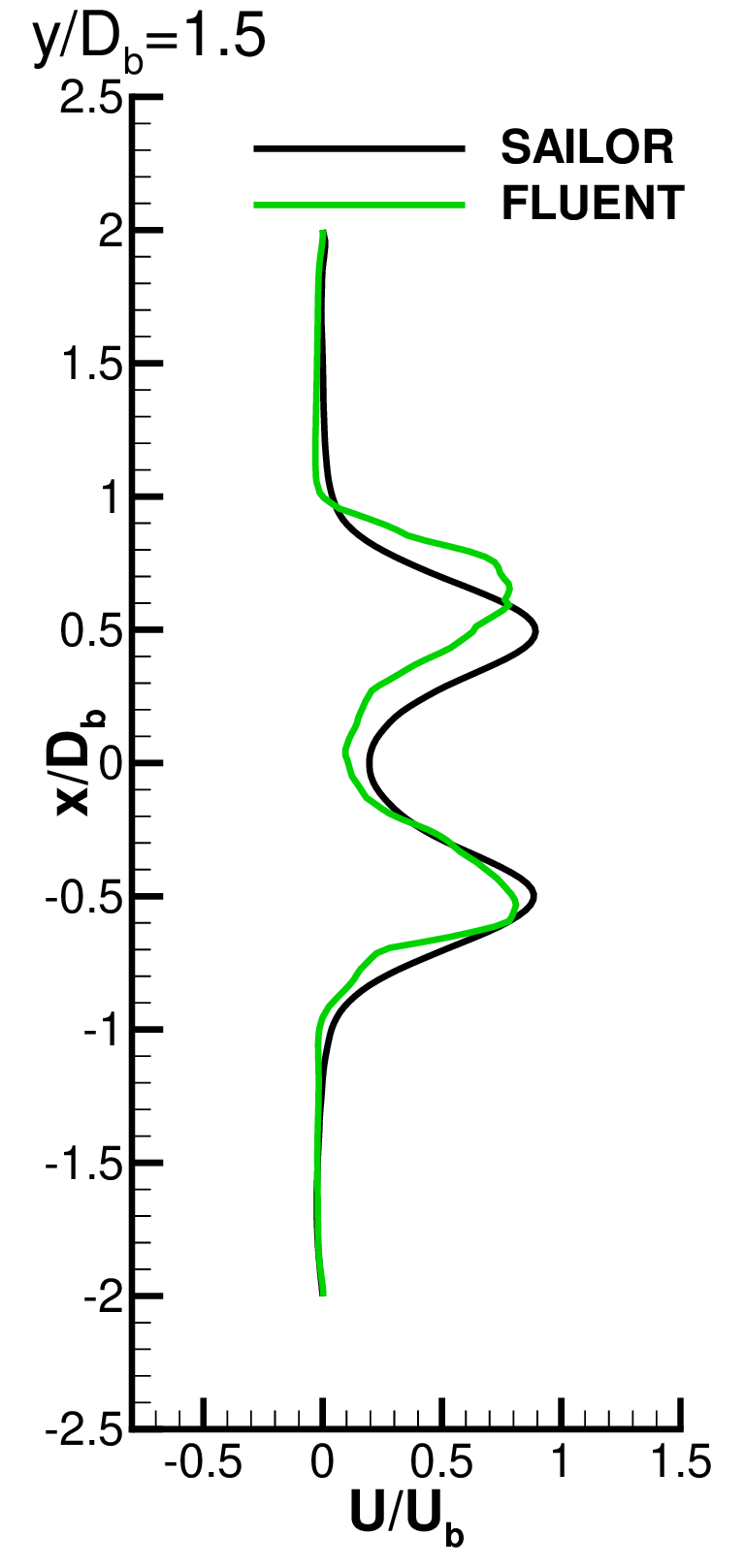}
\end{center}
	\caption{Axial velocity profiles at three different heights above the bluff-body obtained using Fluent and SAILOR codes.}
	\label{fig:Fluent_SAILOR}
\end{figure}

\begin{figure}[!ht]
 \begin{center}
 \subfigure[cylindrical bluff-body with the flat wall]{
\includegraphics[width=0.45\textwidth]{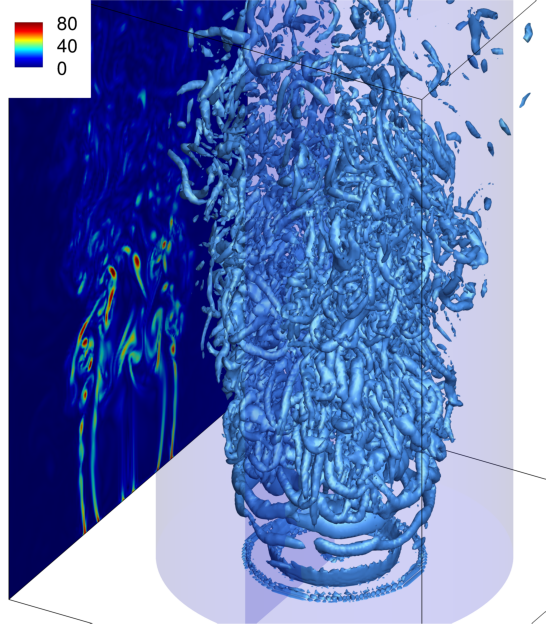}}
 \subfigure[square bluff-body]{
\includegraphics[width=0.45\textwidth]{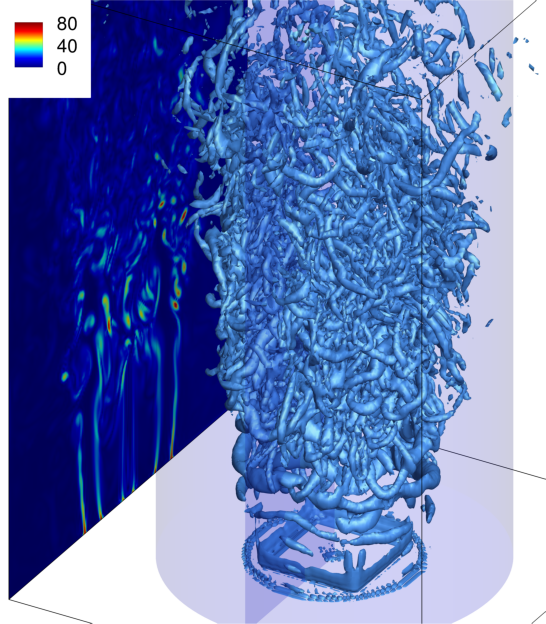}}
 \subfigure[star bluff-body]{
\includegraphics[width=0.45\textwidth]{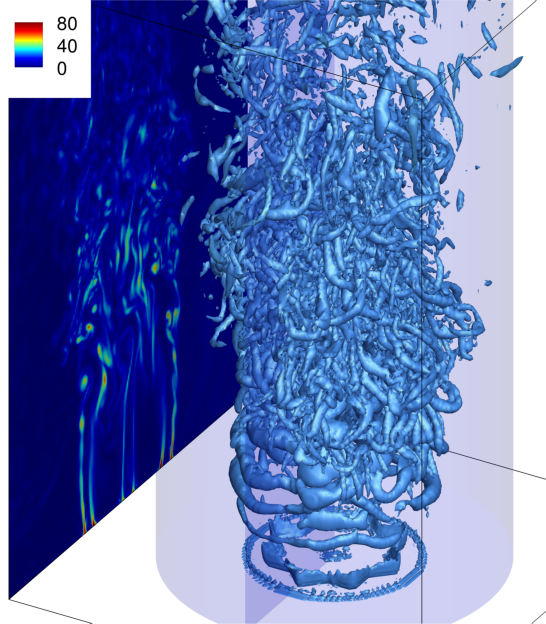}}
 \subfigure[cylindrical bluff-body with the wavy wall]{
\includegraphics[width=0.45\textwidth]{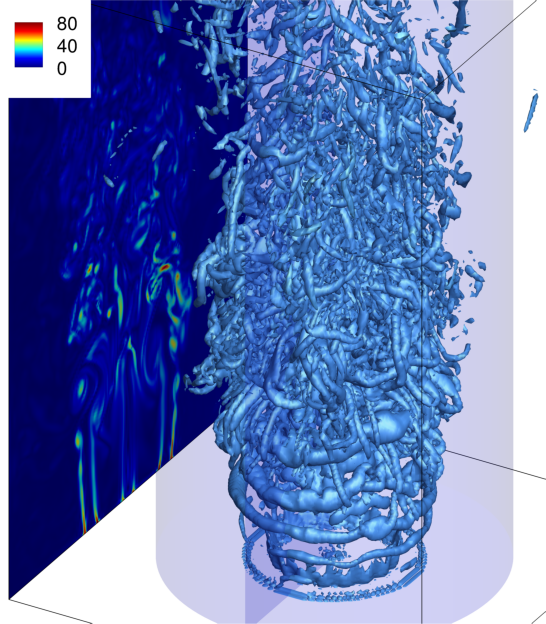}}
\end{center}
\caption{Instantaneous $Q$-parameter isosurfaces and vorticity modulus contours in the central cross-section shown at the boundaries. The partially translucent blue cylindrical surfaces denote a border of subdomain used for the  calculation of the entrainment.}
\label{fig:Q-param}
	\end{figure}

\begin{figure}[!ht]
 \begin{center}
\includegraphics[width=\textwidth]{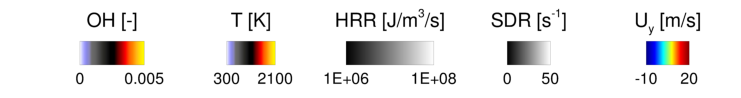}
\includegraphics[width=\textwidth]{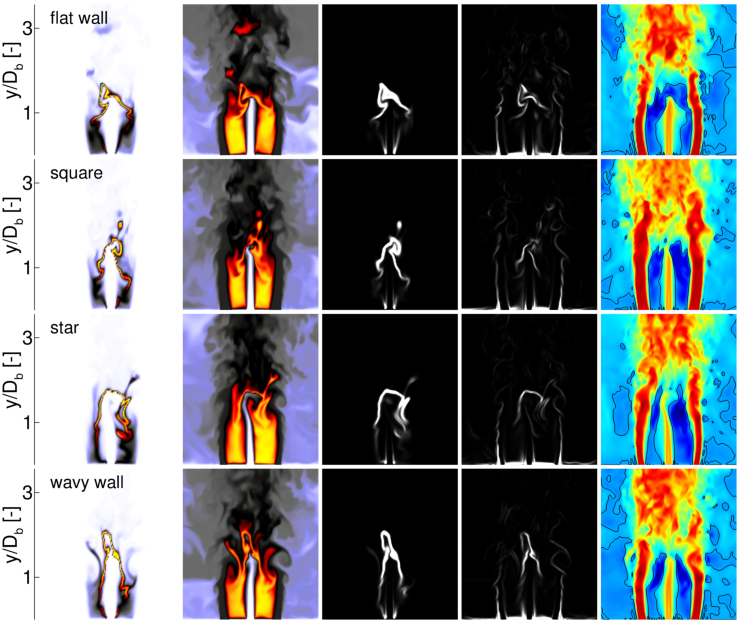}
\includegraphics[width=\textwidth]{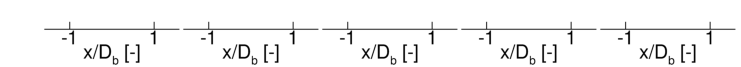}
\end{center}
\caption{Instantaneous distributions of the OH mass fraction, temperature, heat release rate (HRR), scalar dissipation rate (SDR), and the axial velocity. The black line represents its zero level. The results obtained for the various bluff-bodies are shown in the rows.}
\label{fig:inst-results}
	\end{figure}

\subsection{Test case}
In the present work we consider a typical combustion chamber with a conical bluff-body similar to the one studied experimentally by~\cite{kypraiou2018response}. We replace the methane by the hydrogen diluted with nitrogen, which is an alternative zero-carbon fuel. The hydrogen mass fraction in the fuel stream is equal to 0.05. A schematic view of the analysed configuration is presented in Fig.~\ref{fig:BB-config}. The cold  fuel (300 K) is injected to the chamber through the 4 mm slot in a fuel pipe ended with a bluff-body surface. The bulk axial velocity of the fuel is assumed equal to $10$~m/s. Inside the chamber the fuel ignites in a co-flowing air heated to 1000~K. A global equivalence ratio is equal to 0.047 and the theoretical power generated by the combustion process is 0.5 kW.
Unlike in the original configuration of~\cite{kypraiou2018response} we do not add the swirl to the oxidizer stream and the dominant effect on the flame is due to the changing geometry of the bluff-body. \\

Concerning the dynamics of the bluff-body stabilized flames, directly above the bluff-body a central recirculation zone (CRZ in Fig.~\ref{fig:BB-config}) is formed. Its dimensions and inner structure depend on the bluff-body size and flow parameters. Inside CRZ a smaller inner recirculation zones (IRZs) may exist.
The bluff-bodies investigated in this paper are characterised by the equivalent diameter $D_b=2\,\sqrt[]{S/\pi}$=25 mm ($S$ - actual area of bluff body) and they are placed in a circular duct of diameter $D$=37 mm. Four different bluff-bodies are considered, the cylindrical with a flat and wavy wall (waviness oriented streamwise), square, and hexagram (star) one. All geometries are displayed in Fig.~\ref{fig:BB-config} along with the computational meshes prepared in the ANSYS Meshing module.

\subsection{Numerical details}
As mentioned before, the simulations are performed using a two-stage approach schematically presented in Fig.~\ref{fig:BB-config}. In the first stage (I) the ANSYS Fluent solver is used to model the flow through the entrance duct and around the bluff-bodies. During these calculations we acquire unsteady velocity signals at the end of the inlet section for a period of 150$D_b/U_b$ ($U_b$- bulk velocity of the co-flow).
The computational grids for all bluff-bodies generated in ANSYS Meshing module are shown in the central cross-section in Fig.~\ref{fig:BB-config}. They are block-structured and precisely fitted to the shape of each bluff-body. All geometries are discretised in this way that a near-wall cell size allows for a proper resolution of the turbulent boundary layers ($y+\approx 1$). The minimum element size is assumed at the level of 0.5 mm for the bluff-body with the flat wall, whereas in the case of the wavy-wall a finer mesh  in the near-wall region is used with the cell height 0.25 mm. The overall numbers of cells in the inlet duct are equal to 3.6$\times 10^5$ and 3.7$\times 10^5$ for the cylindrical bluff-body with the smooth and wavy wall, respectively.\\

In the second part of the calculations (stage II) the SAILOR code is applied, which so far has been used in a number of studies devoted to combustion problems in jet type flames~\citep{Tyliszczak_CNF_2015,rosiak2016cmc} and mixing layers~\citep{WarzakTyliszczak_CNF_2019,WawrzakTyliszczak_FTAC_2020}.
The extracted velocity fields from the stage I of the simulations are imposed onto the inlet plane of the computational domain involving only the outer section of the burner (see Fig.~\ref{fig:BB-config}). There is no heat transfer between the flame and the bluff-body, since the bluff-body is only embedded in the preliminary Fluent simulations. In a real situation, a flame attached to a bluff-body would certainly lose heat toward it. However, we neglected this effect and conducted comparative studies focusing on flow structure affected by different bluff-bodies. One would expect that the possible errors resulting from the applied procedure would be similar in all cases.

In stage II, the computational domain extends 6$D_b$ in the axial direction and 4$D_b$ in the radial and tangential directions. A grid independent solution is obtained at a grid size $N_x\times N_z\times N_y=144\times 144 \times 192$ nodes. The nodes are compacted axially and radially but almost a uniform mesh is constructed close to the inlet section. The maximum cell volume is equal to $1.7\times 10^{-9}$ m$^3$, that corresponds to the minimum cell volume of the computational meshes used in the ANSYS software. The time-step is computed according to the Courant-Friedrichs-Lewy (CFL) condition with the CFL number assumed equal to 0.25.

Concerning the boundary conditions, at the side boundaries the velocity is set to zero and the species and enthalpy are computed form the Neumann condition, i.e., $\partial {Y}_{\alpha}/\partial n=0$ and $\partial h/\partial n=0$. The pressure is computed from the Neumann condition ($\partial p/\partial n=0$) both at the side boundaries as well as at the inlet plane. At the outlet boundary, all velocity components, species and enthalpy are computed from the convective boundary condition $\partial C /\partial t + V_c \partial C/\partial y=0$, where $C$ represents a general variable and $V_c$ is the convection velocity calculated as the mean velocity in the outlet plane. To avoid back-flow the velocity $V_c$ is limited such that $V_c=\max(V_c,0)$. The pressure at the outflow is assumed constant and equal to $p=101325$~Pa.

The SAILOR code uses the projection method for pressure-velocity coupling~\citep{Tyl16} combined with a predictor-corrector method (Adams-Bashforth / Adams-Moulton) applied for the time integration. The spatial discretization is performed on half-staggered meshes applying the 6th order compact finite difference approximation for the momentum and continuity equations. The second-order TVD (Total Variation Diminishing) scheme with Koren limiter is used for the transport equations for chemical species and enthalpy. The time integration of the chemical source is performed with the help of the {VODPK} (Variable-coefficient Ordinary Differential equation solver with the Preconditioned Krylov method) solver~\citep{VODPK} that is well suited for stiff systems. The chemical reaction terms are calculated using the {CHEMKIN} interpreter and the chemical reactions are computed using detailed mechanism of Mueller et al.~\citep{Muetal99} involving 9 species and 21 reactions.\\

Preliminary simulations performed using ANSYS Fluent and the SAILOR code for the air stream flowing in the duct around the cylindrical bluff-body were conducted to verify the solution strategy. The comparison of the velocity profiles behind the bluff-body revealed that an interpolation of the velocity components required for the two-stage computational procedure does not introduce significant errors. As can be seen in Fig.~\ref{fig:Fluent_SAILOR} both solvers provided similar velocity evolution downstream of the bluff-body. However, the results obtained using the SAILOR code seem to be more accurate, at least qualitatively, as they reflect the symmetric shape of the bluff-body. The applied two-stage approach was also proven to yield the correct results in previous studies devoted to passively controlled jet flames issuing from polygonal nozzles~\citep{KubanStempkTyliszczak_Energies_2021}.
\begin{figure}[!ht]
 \begin{center}
\subfigure[cylindrical bluff-body with the flat wall]{
\includegraphics[width=0.45\textwidth]{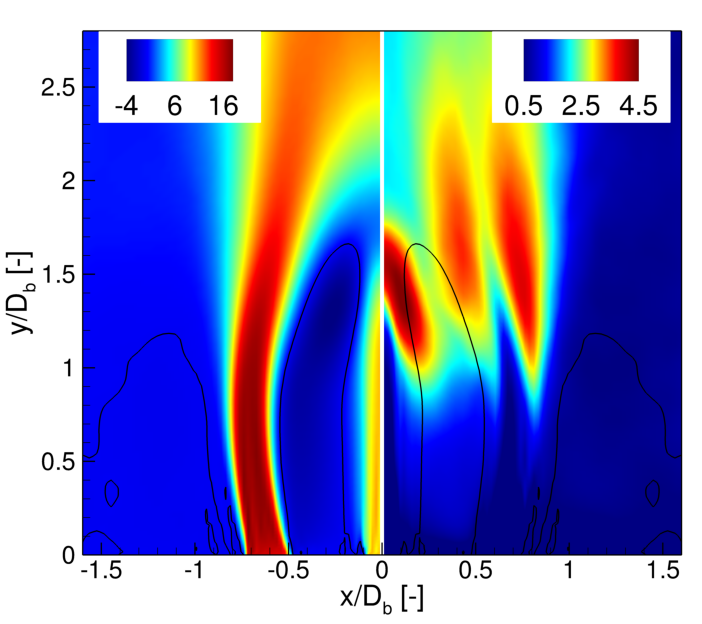}}
\subfigure[square bluff-body]{
\includegraphics[width=0.45\textwidth]{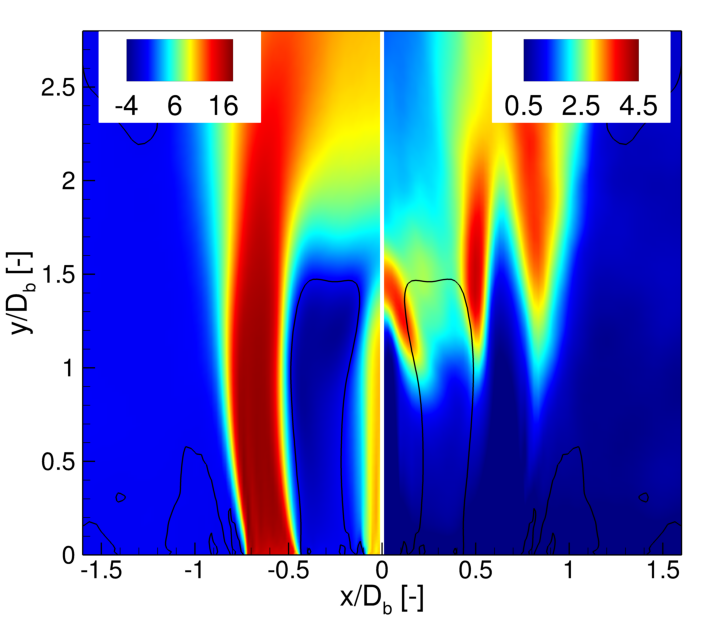}}
\subfigure[star bluff-body]{
\includegraphics[width=0.45\textwidth]{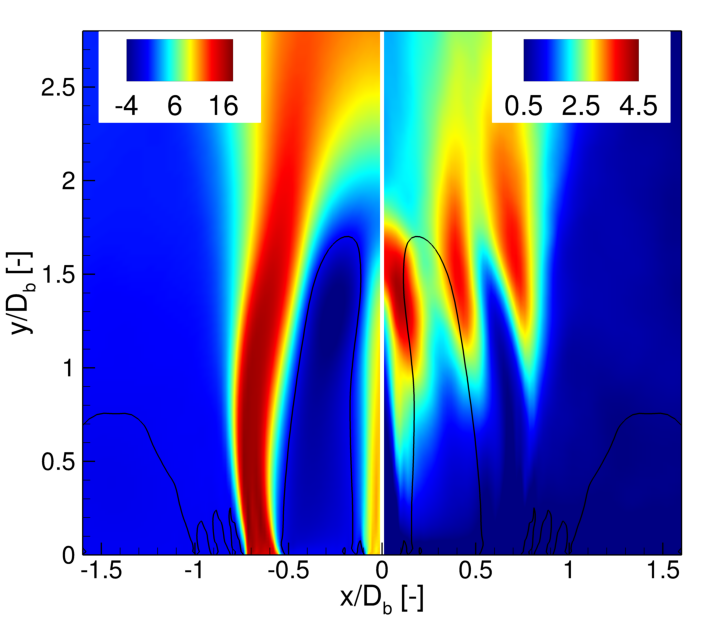}\label{subfig:Vmean_rms_star}}
\subfigure[cylindrical bluff-body with the wavy wall]{
\includegraphics[width=0.45\textwidth]{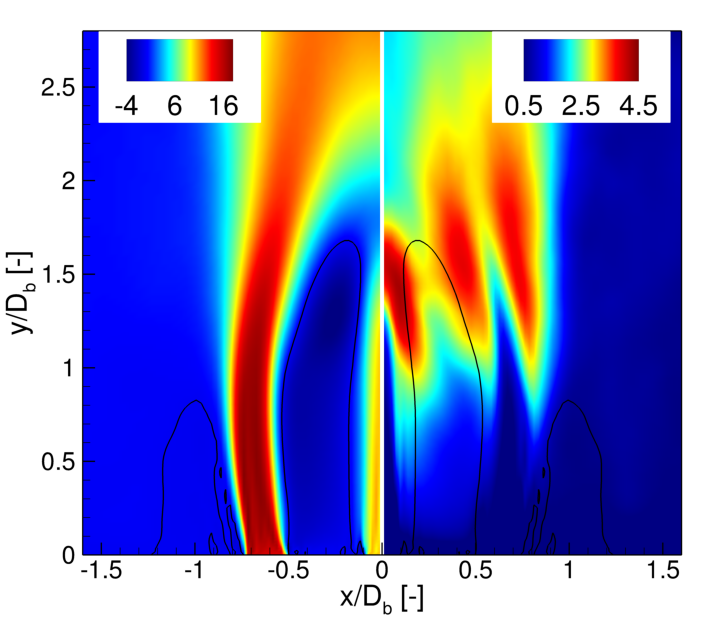}}
\end{center}
\caption{Time-averaged axial velocity contours (mean-left, r.m.s-right) and isolines of the zero axial velocity ($\left< U_y\right>=0$).}
\label{fig:Vmean_rms}
	\end{figure}
%
\section{Results}
\subsection{Instantaneous flame dynamics}
As the streams of the fuel and oxidizer are injected from the inlet plane into a quiescent ambient fluid the flow undergoes the Kelvin-Helmholtz instability due to velocity gradients in the shear layers. The vortex rings are generated and detached from the bluff body edge. Their shape is respective to the bluff-body shape that can be deduced based on the $Q$-parameter isosurfaces presented in Fig.~\ref{fig:Q-param}. In this figure, the blue cylinders denote the control surface used to predict the entrainment rate, which will be discussed later. Additionally, the vorticity magnitude plotted on the central cross-section is shown at the sides.
Depending on the bluff-body, the flow fields differ and it can be seen that with increasing bluff-body complexity the flow patterns become  more complex. Since the initial vortices are considerably dependent on the shape and wall topology of the bluff-body, their distortion changes as they travel downstream. Regarding the circular bluff-body, an earlier formation of streamwise vortical structures takes place for the case with the flat wall, whereas stronger vortex rings are observed  when the wall is wavy. The results obtained for the cases with the square and star bluff-bodies reveal very complex flow structure. The presence of small-scale vortices generated behind the sharp corners causes a deformation of large vortical structures closer to the inlet compared to the case with the circular bluff-body.
Comparison of the vorticity contours reveals that flow inside CRZ is the most energetic for the cases with the non-circular shapes, consistently with the findings for non-reacting jet flows (e.g.~\cite{tyliszczak2022numerical}). For the cases with the square and star bluff-bodies the vorticity contours seem to be more wrinkled, especially in the region closer to the inlet in the outer shear layer. Keeping in mind that the flames reflect the shapes of the bluff-bodies close to the inlet plane, it is expected that the impact of large-scale intense vortices  on  the flame structure and its shape is prominent.\\
\begin{figure}[t]
 \begin{center}
\subfigure[mean axial velocity]{
\includegraphics[width=0.45\textwidth]{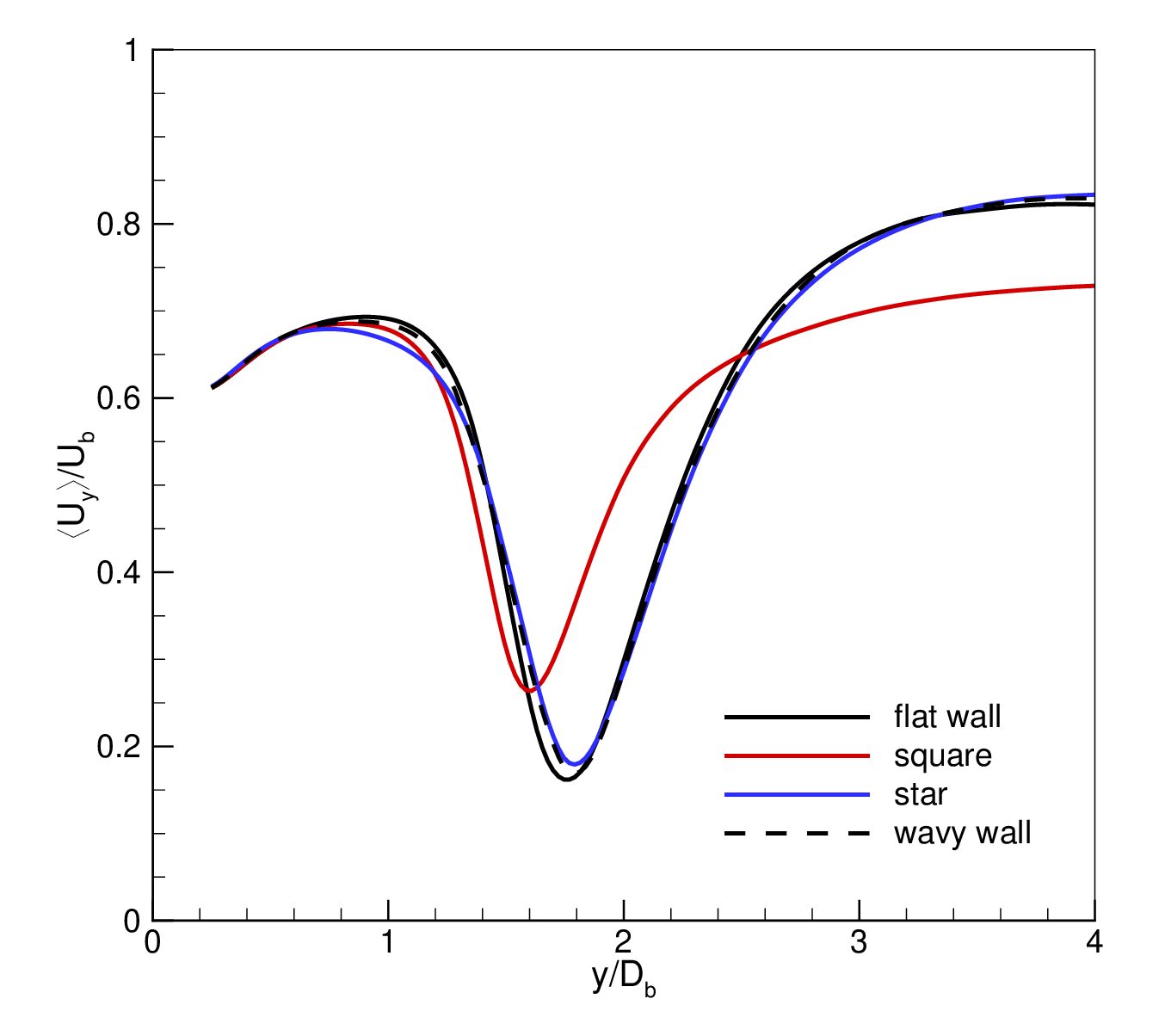}\label{subfig:axial_Vmean}}
\subfigure[rms axial velocity]{
\includegraphics[width=0.45\textwidth]{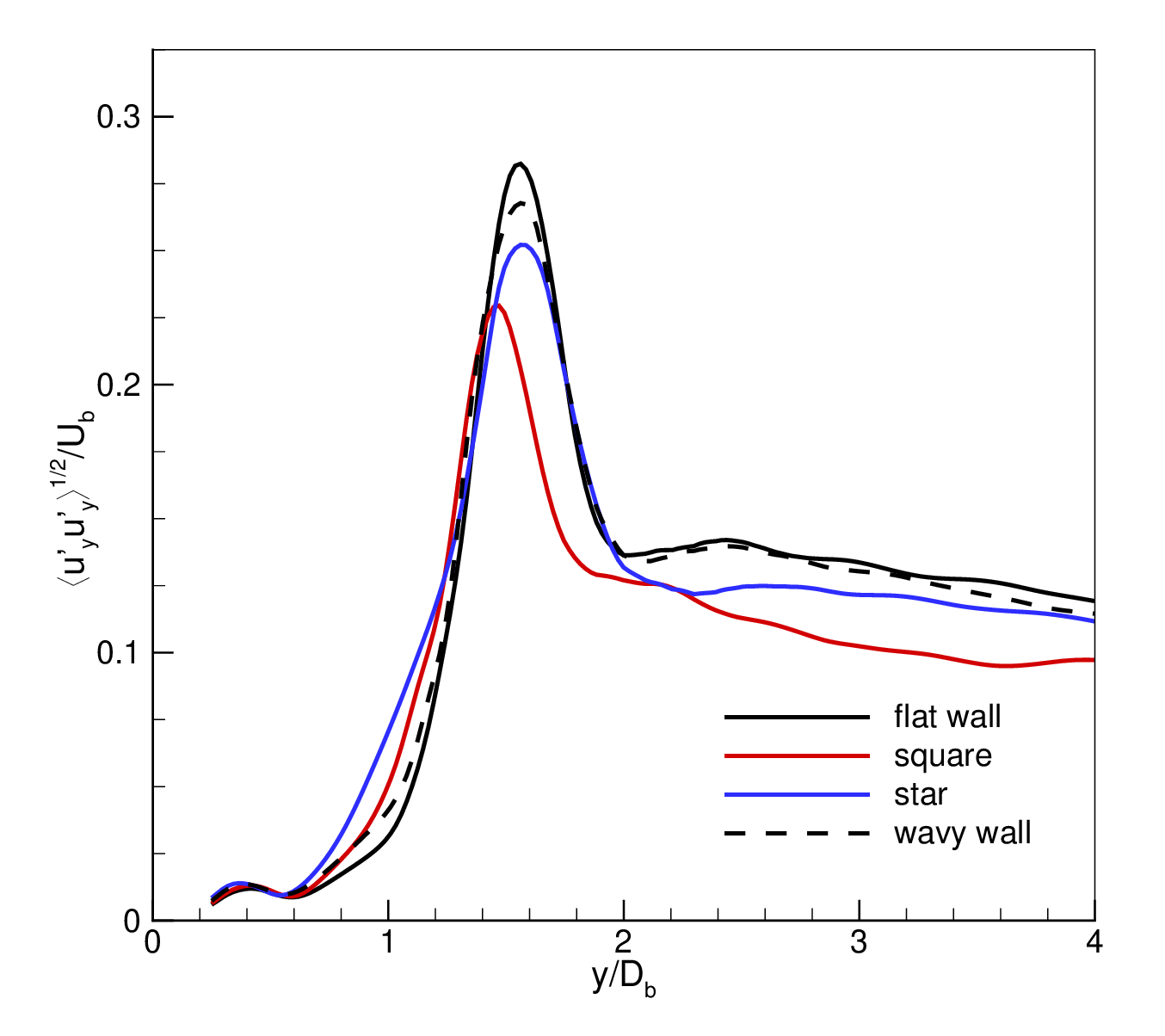}\label{subfig:axial_Vrms}}
\subfigure[entrainment rate]{
\includegraphics[width=0.45\textwidth]{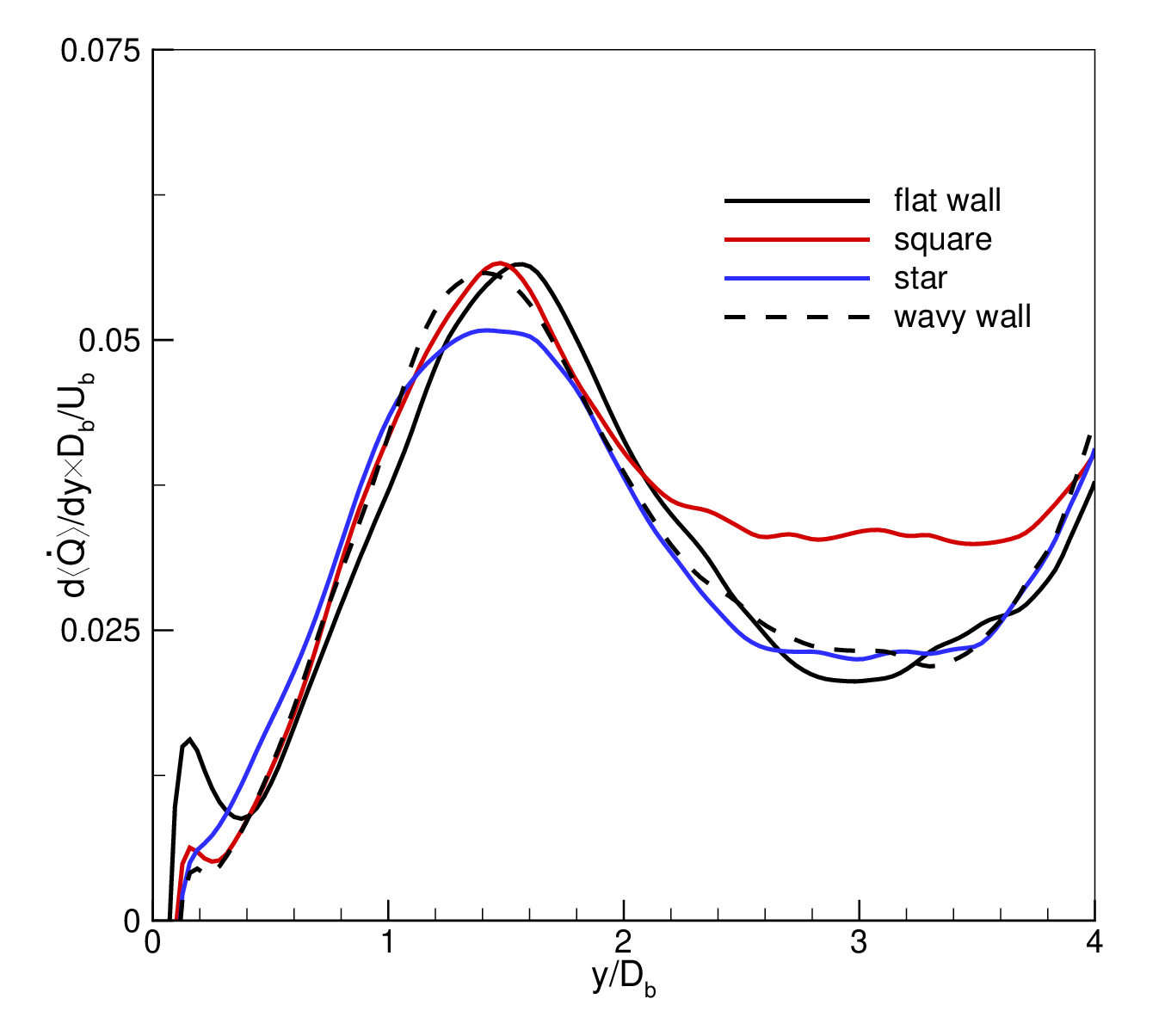}\label{subfig:entrainment}}
\end{center}
\caption{Centerline profiles of the time-averaged axial velocity (a) and its fluctuations (b) and the profiles of the entrainment (c).}
\label{fig:axial_V}
	\end{figure}

Figure~\ref{fig:inst-results} illustrates the instantaneous distributions of OH species mass fraction, temperature, heat release rate (HRR), scalar dissipation rate (SDR), and the axial velocity. It can be observed that the flames are stabilized by the reversed flow (see the black isolines indicating the recirculation zones) that also manifests by a large level of OH species in the inner shear layer. The heat release rate is found to be the strongest near the flame edge where also the OH species is abundantly produced. Both the OH and HRR distributions indicate relatively low reactive regions inside the recirculation zones within the axial distance $y/D=0.4-0.8$. Here, the fuel is rapidly mixed with the recirculating hot air and combustion products promoting the partial oxidation accompanied by a small amount of heat release. The scalar dissipation rate is low in these regions, contrary to the vicinity of the bluff-body flame and downstream axial locations.
Finally, it can be seen that the recirculation zone, which have the shape depending on bluff-body, controls the shape of the flame. 

\subsection{Statistical properties}
\begin{figure}[!ht]
 \begin{center}
\subfigure[mean H$_2$]{
\includegraphics[width=0.45\textwidth]{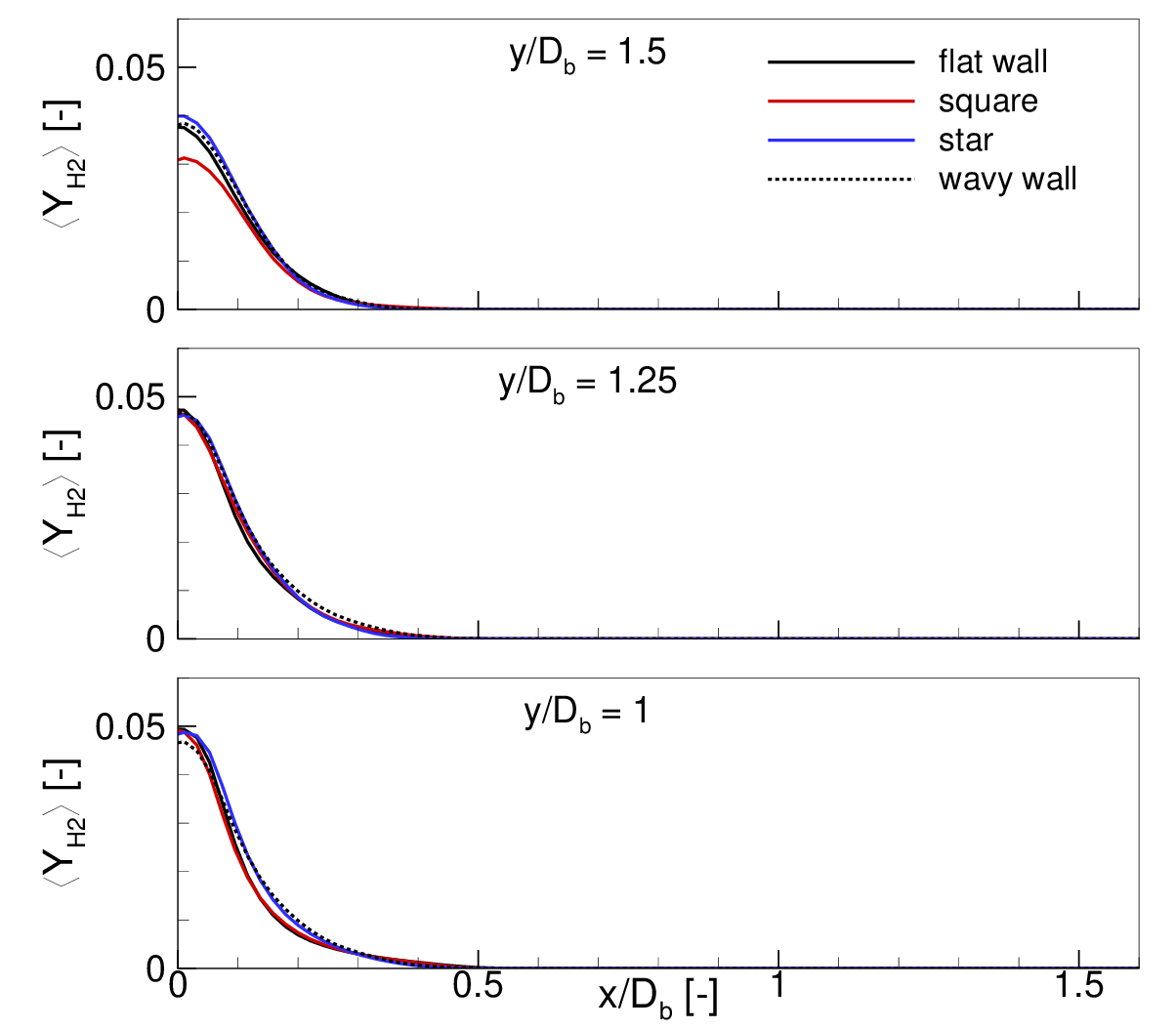}\label{subfig:YH2_radial}}
\subfigure[mean temperature]{
\includegraphics[width=0.45\textwidth]{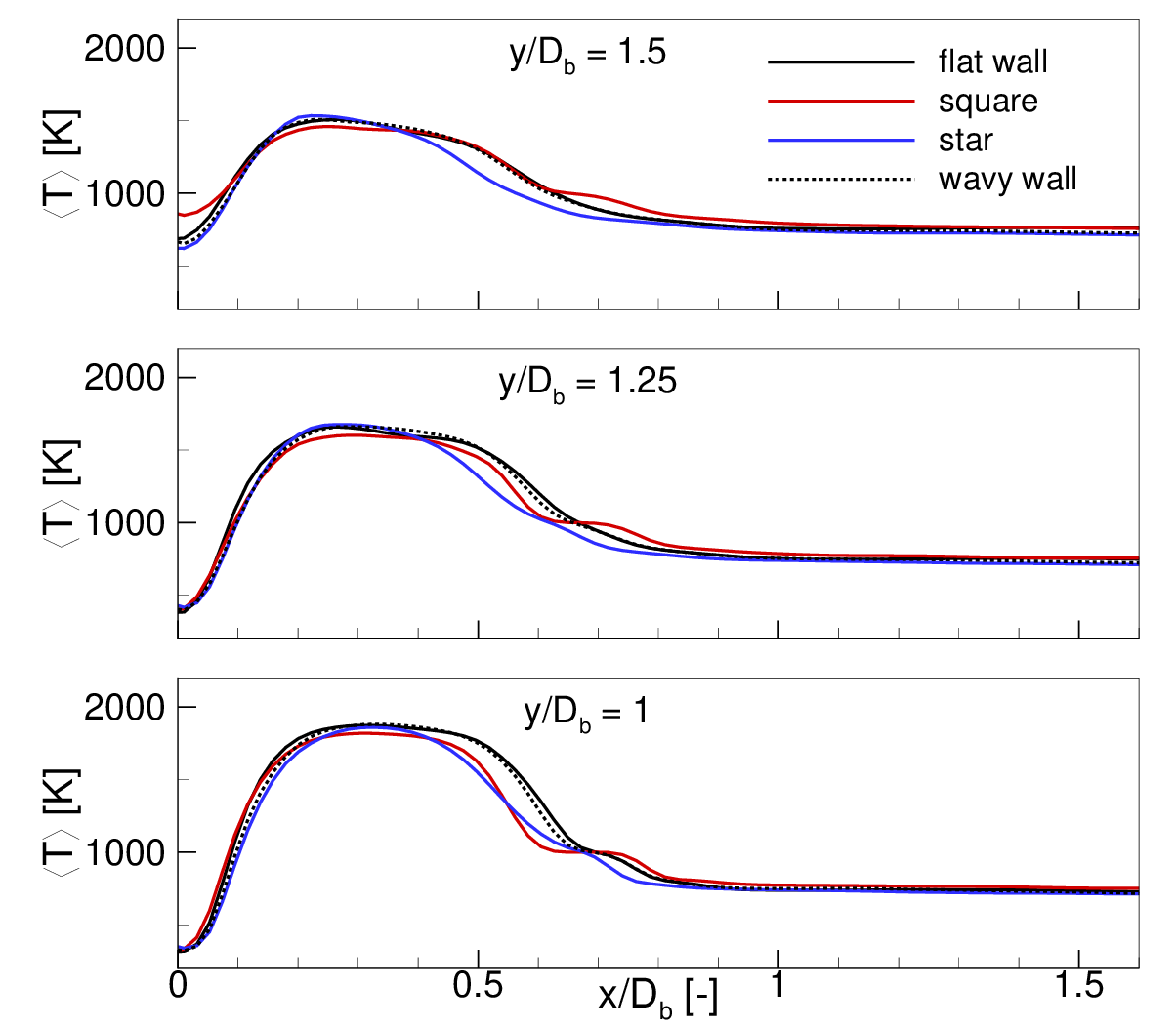}}
\subfigure[rms temperature]{
\includegraphics[width=0.45\textwidth]{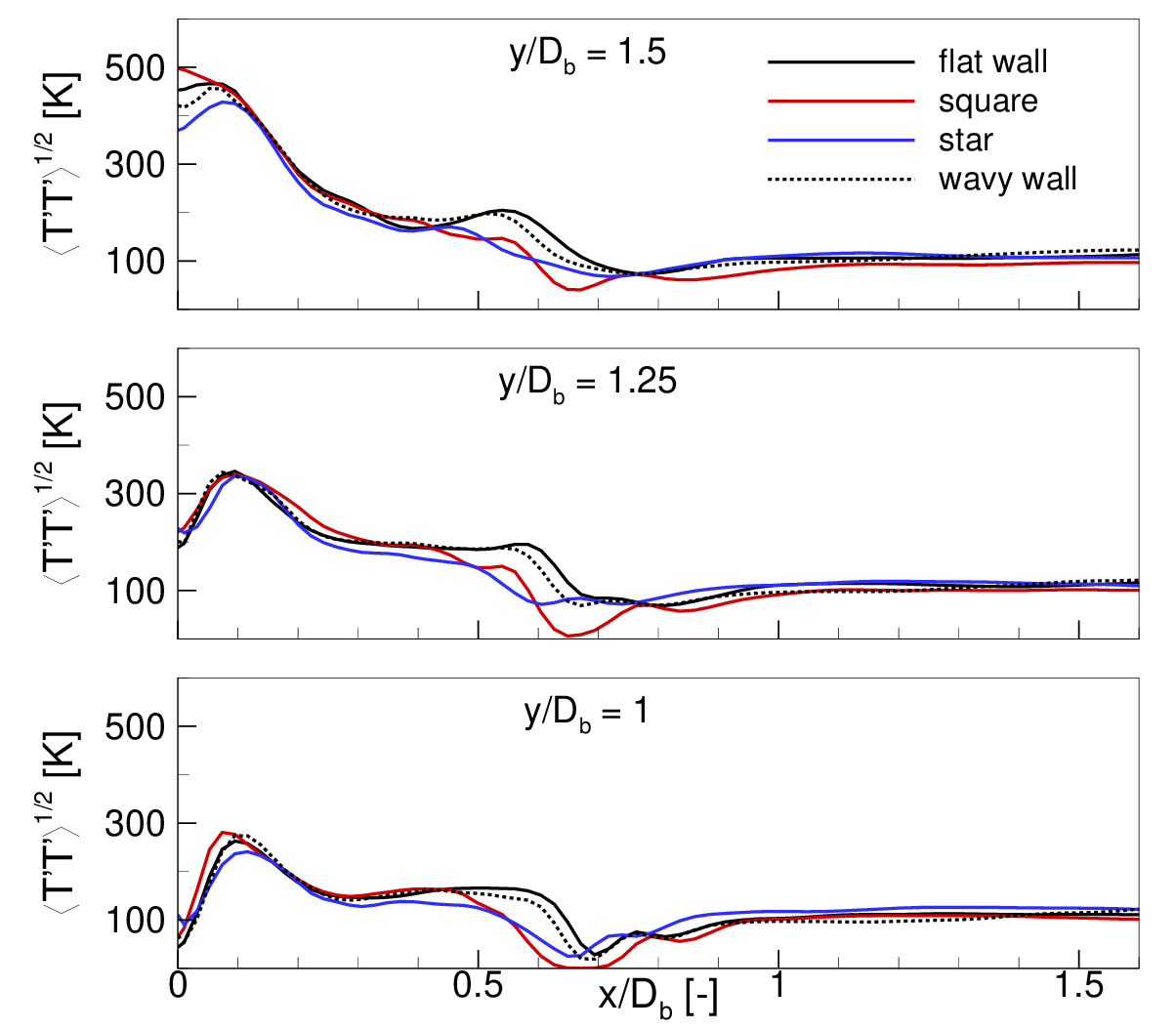}\label{subfig:Trms_radial}}
\end{center}
\caption{Radial profiles of the time averaged H2 species mass fraction and temperature (mean and rms) at different axial locations.}
\label{fig:radial_prof}
	\end{figure}

Detailed analysis of the flame dynamics in CRZ has been carried out based on the time-averaged results presented in Figs.~\ref{fig:Vmean_rms}-\ref{fig:radial_prof}. The time-averaging procedure started after the flames fully developed in the domain and was continued for 250$D_b/U_b$ time units resulting in well-converged statistical results.
Figure~\ref{fig:Vmean_rms} presents the contours of the time-averaged axial velocity and its fluctuations (r.m.s) for all cases considered. The location of the recirculation zone is represented by the isolines of $\left< U_y\right>=0$. As can be seen, the narrowing of the central recirculation zones is caused by the  the oxidizer stream. This is accompanied by a significant increase of the axial velocity fluctuations due to a strong shear forces.
It is evident that the location, size and shape of the recirculation zones are only slightly affected by the wall topology compared to the changes caused by the bluff-body shape.
However, there are noticeable differences regarding the cases with the flat and wavy walls. 
Unlike as in the case with the flat wall, the isoline of the zero axial velocity in the case with the wavy wall is nearly parallel to the fuel stream. It results in an enhanced mixing that manifests by a more uniform velocity field inside CRZ. 
It is also clearly seen that the recirculation zone shrinks radially in the case with star bluff-body. It significantly changes the shape also when the square bluff-body is used. In this case, the fluctuation maximum is shifted beyond the recirculation zone.\\

Centerline profiles of the mean and r.m.s axial velocity normalized by $U_b$ are presented in Figs.~\ref{subfig:axial_Vmean} and~\ref{subfig:axial_Vrms}. In the initial part of the flow they represent mainly the velocity of the fuel. It can be seen that in all the cases it starts to decay almost in the same axial locations. The smallest drop is found for the case with the square bluff-body. In this case, the axial velocity minimum is about $U_{y}/U_{b}$=0.1 larger and is shifted $15\%D_b$ towards the bluff-body. Concerning the differences in the fluctuations profiles it can be seen that for the case with the square bluff-body the fluctuations maximum occurs earlier (0.1$D_b$) and reaches approximately 5\% lower level compared to the remaining cases. 
The shift and smaller value of the velocity minimum are the result of intensified mixing caused by the small vortices generated at the sharp corners. This suggests that the square bluff-body can ensure the most intensified fuel/oxidizer mixing.\\

The entrainment is calculated following~\cite{wygnanski1969some} as:
\begin{equation}
\frac{d\dot{Q}}{dy}=\ointop_{c}\mathbf{U}\cdot\mathbf{n}dc=\int_{0}^{2\pi}\left\langle V_r\right\rangle r d\theta
\end{equation} 
where $\dot{Q}$ is the volume flux, $c$ is a path outside a given region, $\mathbf{U}$ and $\mathbf{n}$ are the velocity and normal vectors, $V_r$ is the radial velocity component. In this work we calculate the entrainment over a surface of a cylindrical domain with the radius 1.5$D_b$ that was shown in Fig.~\ref{fig:Q-param}. This cylinder encloses the flow region in which the majority  large vortical structures fits.

The entrainment rate profiles presented in Fig.~\ref{subfig:entrainment} shows that starting from $y/D>0.2$ the volume flux linearly increases up to $y/D\approx 1.5$ that is approximately the distance where the recirculation zones end. It drops further downstream and reaches the local minimum at $y/D\approx 3$. Concerning the impact of the bluff-body geometry on the entrainment, it can be seen that only minor differences between the cases with the flat and wavy exist. The most pronounced one seems to be the faster entrainment growth immediately above the bluff-body surface for the case with the flat wall. The situation changes when the star bluff-body is used. In this case, starting from approximately $y/D=0.1$ the entrainment increases the fastest. This is caused by the vortices generated in the acute corners and their quicker break-up. They also alter the central flow downstream and therefore the maximum occurred at $y/D\approx 1.5$ is lower than in the remaining cases. Regarding the region above the recirculation zones, i.e., $y/D>2$, the solution obtained for the configuration with the square bluff-body differs the most. In this case, rather than the local minimum the profile reveals a plateau at relatively high level. A similar shape of the entrainment in this region is observed for the star bluff-body, yet, in this case the values of the entrainment are more or less the same as in the cases with the cylindrical bluff-bodies.\\
The increase of the entrainment in the configurations with the star and square bluff-bodies influences the radial distribution of the hydrogen mass fraction and temperature. Their profiles in three different axial distances inside the recirculation zone ($x/D_b$=1, $y/D_b$=1.25, $y/D_b$=1.5) are presented in Fig.~\ref{fig:radial_prof}. The enhancement of the mixing for the case with the square bluff-body is manifested by a reduced fuel mass fraction in the axis at the  distance $y/D_b$=1.5 (see also Fig.~\ref{subfig:YH2_radial}). Moreover, for this case the  temperature  at $y/D_b$=1.5 is approximately 200~K higher compared to the remaining cases. This means a more intense combustion process in the central part of the flow. On the other hand, in the region of the shear layer $x/D_b$=0.5 the lowest temperature is found in the configuration with the star bluff-body, especially at $y/D_b$=1.25 and $y/D_b$= 1.5. Apparently, in the configuration with the star shape bluff-body the flame is noticeably shifted towards the axis, whereas in the case with the square bluff body it seems compacted towards the inlet plane. This is in line with the observation made for the velocity field and CRZ structure. \\

Radial profiles of the temperature fluctuations presented in Fig.~\ref{subfig:Trms_radial} also show important differences between particular solutions. Initially ($y/D_b$=1-1.25), the local maximum of the temperature fluctuations persists in the inner shear layer at $x/D_b$=0.1 for all the cases. Then, the fluctuations consistently rise and the maximum shifts towards the axis. It appears first on the axis at $y/D_b$=1.5 when the square bluff-body is used. In this case, the contribution of the fluctuating component is considerably higher close to the inlet plane and is transferred further downstream with 40 K higher values than for the cylindrical bluff-body with flat wall. Such increase in the fluctuating quantities consistently affects the mean component. 
Moreover, it can be seen that there are no substantial differences in the profiles obtained for different wall topologies.
\section{Summary}
The paper presented the analysis of flame structures behind a conical bluff-body burner having different wall topologies and shapes. The research was performed using the LES method and two numerical tools, ANSYS Fluent code and in-house code SAILOR. The former allowed for accurate simulations of the flow around the complex shape bluff-bodies and the latter for precise modelling of the combustion process.  
It has been shown that in the vicinity of the bluff-body its shape has a strong influence on vortices induced in the shear layer formed between the central recirculation zone and the oxidizer stream.  Compared to the conventional circular shape, the acute corners caused the enhancement of the mixing processes in the central recirculation zone.  
The impact of change in geometry shape and wall topology observed in instantaneous results were quantified based on the time-averaged plots of mean and fluctuating velocity, fuel mass fraction and temperature. In general, it has been observed that in the configurations with the star and square bluff-bodies the flame in the recirculation zone is shifted towards the injected fuel stream. 
A faster oxidizer entrainment observed in these cases caused an enhanced mixing process. The analysis of the velocity distributions revealed a shortened recirculation zone by 15\% of the equivalent bluff-body diameter and the smallest axial velocity drop for the configuration with the square bluff-body. This resulted in a significantly larger temperature in the centerline of the flame and higher temperature fluctuations. The influence of the wall topology (flat vs. wavy) in the configuration with the classical conical bluff-body turned out to be very small. It resulted only in minor modifications of the flow structures in the direct vicinity of the inlet plane but this had practically no impact on the flame formed downstream. 

\section*{Acknowledgment}
This work was supported by the National Science Center in Poland (Grant 2020/39/B/ST8/02802) and statutory founds (BS/PB-1-100-3011/2022/P). The computations were carried out using the PL-Grid Infrastructure.



\begin{thebibliography}{19}
\providecommand{\natexlab}[1]{#1}
\providecommand{\url}[1]{\texttt{#1}}
\expandafter\ifx\csname urlstyle\endcsname\relax
  \providecommand{\doi}[1]{doi: #1}\else
  \providecommand{\doi}{doi: \begingroup \urlstyle{rm}\Url}\fi

\bibitem[Docquier and Candel(2002)]{docquier2002combustion}
N.~Docquier and S.~Candel.
\newblock Combustion control and sensors: a review.
\newblock \emph{Progress in Energy and Combustion Science}, 28\penalty0
  (2):\penalty0 107--150, 2002.

\bibitem[Dr{\'o}{\.z}d{\.z} et~al.(2021)Dr{\'o}{\.z}d{\.z}, Niegodajew,
  Roma{\'n}czyk, Sokolenko, and Elsner]{drozdz2021effective}
A.~Dr{\'o}{\.z}d{\.z}, P.~Niegodajew, M.~Roma{\'n}czyk, V.~Sokolenko, and W.~Elsner.
\newblock Effective use of the streamwise waviness in the control of turbulent
  separation.
\newblock \emph{Experimental Thermal and Fluid Science}, 121:\penalty0 110291,
  2021.

\bibitem[Duwig et~al.(2011)Duwig, Nogenmyr, Chan, and Dunn]{duwig2011large}
C.~Duwig, K.~Nogenmyr, C.~Chan, and M.~J.~Dunn.
\newblock Large eddy simulations of a piloted lean premix jet flame using
  finite-rate chemistry.
\newblock \emph{Combustion Theory and Modelling}, 15\penalty0 (4):\penalty0
  537--568, 2011.

\bibitem[Geurts(2004)]{geurts2004elements}
B.~Geurts.
\newblock \emph{Elements of Direct and Large-eddy Simulation}.
\newblock R.T. Edwards, 2004.

\bibitem[Jones and Navarro-Martinez(2007)]{jones2007large}
W.~P.~Jones and S.~Navarro-Martinez.
\newblock Large eddy simulation of autoignition with a subgrid probability
  density function method.
\newblock \emph{Combustion and Flame}, 150\penalty0 (3):\penalty0 170--187,
  2007.

\bibitem[Kuban et~al.(2021)Kuban, Stempka, and
  Tyliszczak]{KubanStempkTyliszczak_Energies_2021}
L.~Kuban, J.~Stempka, and A.~Tyliszczak.
\newblock Numerical analysis of the combustion dynamics of passively controlled
  jets issuing from polygonal nozzles.
\newblock \emph{Energies}, 14\penalty0 (3):\penalty0 554, 2021.

\bibitem[Kypraiou et~al.(2018)Kypraiou, Allison, Giusti, and
  Mastorakos]{kypraiou2018response}
A.~M.~Kypraiou, P.~M. Allison, A.~Giusti, and E.~Mastorakos.
\newblock Response of flames with different degrees of premixedness to acoustic
  oscillations.
\newblock \emph{Combustion Science and Technology}, 190\penalty0 (8):\penalty0
  1426--1441, 2018.

\bibitem[Mi and Nathan(2010)]{mi2010statistical}
J.~Mi and G.~J.~Nathan.
\newblock Statistical properties of turbulent free jets issuing from nine
  differently-shaped nozzles.
\newblock \emph{Flow, turbulence and combustion}, 84\penalty0 (4):\penalty0
  583--606, 2010.

\bibitem[Mueller et~al.(1999)Mueller, Kim, Yetter, and Dryer]{Muetal99}
M.~A. Mueller, T.~J. Kim, R.~A. Yetter, and F.~L. Dryer.
\newblock Flow reactor studies and kinetic modeling of the {H2/O2} reaction.
\newblock \emph{International Journal of Chemical Kinetics}, 31\penalty0
  (2):\penalty0 113--125, 1999.

\bibitem[{P.~N. Brown} and {A.C. Hindmarsh}(1989)]{VODPK}
{P.~N. Brown} and {A.~C. Hindmarsh}.
\newblock {Reduced Storage Matrix Methods in Stiff ODE Systems}.
\newblock \emph{{Journal of Applied Mathematical Computations}}, 31:\penalty0
  40--91, 1989.

\bibitem[Rosiak and Tyliszczak(2016)]{rosiak2016cmc}
A.~Rosiak and A.~Tyliszczak.
\newblock {LES-CMC} simulations of a turbulent hydrogen jet in oxy-combustion
  regimes.
\newblock \emph{International Journal of Hydrogen Energy}, 41\penalty0
  (22):\penalty0 9705--9717, 2016.

\bibitem[Tyliszczak et~al.(2014)Tyliszczak, Cavaliere, and
  Mastorakos]{tyliszczak2014cmc}
A.~Tyliszczak, D.~E. Cavaliere, and E.~Mastorakos.
\newblock {LES}/{CMC} of blow-off in a liquid fueled swirl burner.
\newblock \emph{Flow, Turbulence and Combustion}, 92\penalty0 (1-2):\penalty0
  237--267, 2014.

\bibitem[Tyliszczak(2015)]{Tyliszczak_CNF_2015}
A.~Tyliszczak.
\newblock {LES--CMC} study of an excited hydrogen flame.
\newblock \emph{Combustion and Flame}, 162\penalty0 (10):\penalty0 3864--3883,
  2015.

\bibitem[Tyliszczak(2016)]{Tyl16}
A.~Tyliszczak.
\newblock High-order compact difference algorithm on half-staggered meshes for
  low mach number flows.
\newblock \emph{Computers \& Fluids}, 127:\penalty0 131--145, 2016.

\bibitem[Tyliszczak et~al.(2022)Tyliszczak, Kuban, and
  Stempka]{tyliszczak2022numerical}
A.~Tyliszczak, L.~Kuban, and J.~Stempka.
\newblock Numerical analysis of non-excited and excited jets issuing from
  non-circular nozzles.
\newblock \emph{International Journal of Heat and Fluid Flow}, 94:\penalty0
  108944, 2022.

\bibitem[Vreman(2004)]{Vreman2004}
A.~W. Vreman.
\newblock An eddy-viscosity subgrid-scale model for turbulent shear flow:
  Algebraic theory and applications.
\newblock \emph{Physics of Fluids(1994-present)}, 16\penalty0 (10):\penalty0
  3670--3681, 2004.

\bibitem[Wawrzak and Tyliszczak(2019)]{WarzakTyliszczak_CNF_2019}
A.~Wawrzak and A.~Tyliszczak.
\newblock A spark ignition scenario in a temporally evolving mixing layer.
\newblock \emph{Combustion and Flame}, 209:\penalty0 353--356, 2019.

\bibitem[Wawrzak and Tyliszczak(2020)]{WawrzakTyliszczak_FTAC_2020}
A.~Wawrzak and A.~Tyliszczak.
\newblock Study of a flame kernel evolution in a turbulent mixing layer using
  {LES with a Laminar Chemistry Model}.
\newblock \emph{Flow, Turbulence and Combustion}, 105:\penalty0 807--835, 2020.

\bibitem[Wygnanski and Fiedler(1969)]{wygnanski1969some}
I.~Wygnanski and H.~Fiedler.
\newblock Some measurements in the self-preserving jet.
\newblock \emph{Journal of Fluid Mechanics}, 38\penalty0 (3):\penalty0
  577--612, 1969.

\end{thebibliography}

\end{document}